\documentclass[letterpaper,twocolumn,10pt]{article}
\usepackage{usenix}
\usepackage{graphicx}
\usepackage{caption}
\usepackage{hyperref}
\usepackage{amsmath}
\usepackage{amsfonts}
\usepackage{booktabs}
\usepackage{circuitikz}
\usepackage{subcaption}  % for subfigures
\usepackage{tikz}
\usepackage{svg}
\usepackage{orcidlink}
\svgpath{{images/}}
\usepackage[acronym]{glossaries}

\begin{document}

\date{}
\title{\bf Systematic Security Analysis of the Iridium Satellite Radio Link}
\author{Anonymous Submission}

\author{
{Eric Jedermann\,\orcidlink{0009-0009-0504-5244}}\\
RPTU Kaiserslautern-Landau, Germany\\
jedermann@cs.uni-kl.de
\and
{Piotr Kulpinski}\\
ETH Zurich, Switzerland\\
piotr.kulpinski@outlook.com
%\and
%{Daniele Coppola}\\
%ETH Zurich, Switzerland\\
%dcoppola@ethz.ch 
\and
{\hspace{2cm}Martin Strohmeier\,\orcidlink{0000-0002-1936-0933}}\\
\hspace{2cm}armasuisse, Switzerland\\
\hspace{2cm}martin.strohmeier@armasuisse.ch
\and
{Vincent Lenders\,\orcidlink{0000-0002-2289-3722}}\\
University of Luxembourg, Luxembourg\\
vincent.lenders@uni.lu
\and
{Jens Schmitt}\\
RPTU Kaiserslautern-Landau, Germany\\
jschmitt@cs.uni-kl.de
} % end author

\maketitle

\begin{abstract}

The Iridium Low Earth Orbit (LEO) satellite constellation remains a unique provider of global communications for critical industries, governments, and private users, serving over 2.5 million active subscribers despite recent market competition. In contrast to terrestrial wireless standards such as 3GPP, Iridium’s protocol specifications are proprietary and have not undergone rigorous, public, and systematic security evaluation.
In this work, we present the first comprehensive security analysis of Iridium’s authentication and radio link protocols. We reverse engineer Iridium’s SIM-based authentication mechanism and demonstrate that the secret key can be extracted from the SIM card, enabling full device cloning and impersonation attacks. Leveraging a month-long dataset of Iridium up- and downlink satellite traffic, we further show that nearly all signaling and radio communication protocols currently in use lack encryption, resulting in the exposure of sensitive information in cleartext over the air such as login credentials and large volumes of personal data. Finally, we develop custom software-defined radio (SDR) tools to carry out spoofing and jamming attacks, revealing that modestly equipped adversaries can inject falsified messages or disrupt the Iridium  service locally due to the absence of source authentication.
Our findings uncover systemic vulnerabilities in the Iridium radio link and highlight the urgent need for users of critical applications to transition to more secure communication radio links.

\end{abstract}

\section{Introduction}

Iridium operates a unique Low Earth Orbit (LEO) satellite constellation that delivers global voice and data coverage, including remote and polar regions. With 66 active satellites arranged in six polar orbital planes, Iridium supports critical applications such as maritime and aviation communications, military operations, and Internet-of-Things (IoT) services \cite{iridium-investor-report-2025}. Given the high-stakes nature of these use cases, ensuring the security of Iridium communications is of paramount importance.

Originally designed in the 1990s, Iridium’s network is based on proprietary protocols that have never undergone rigorous, independent, and publicly documented security analysis. This lack of transparency has raised concerns in white hacker circles regarding the confidentiality, integrity, and privacy of the Iridium communications~\cite{pultarova-spectrum2025}. To date, however, no comprehensive and systematic analysis exists for assessing and validating the multitude of security risks that users face when using the Iridium services.

In this work, we aim to fill this gap and present the first systematic security evaluation of the Iridium radio link. Our analysis draws on publicly available developer guides, technical information that has leaked over the years, prior “white-hat” hacker presentations, and our own reverse engineering of Iridium protocols and devices. 

We first define a threat model consisting of six attack primitives targeting the confidentiality, integrity, and availability of user communications over the Iridium radio link. We then experimentally evaluate these primitives in a controlled laboratory environment. Our findings show that Iridium is vulnerable to all six attack types, each of which can be executed over the air using low-cost, off-the-shelf software-defined radio (SDR) equipment.

For example, we observe that Iridium's authentication mechanism is SIM-based and uses a challenge-response protocol analogous to GSM, including the use of the long-deprecated COMP128-1 algorithm for computing authentication responses. COMP128-1 was broken more than two decades ago, enabling recovery of the subscriber secret key $K_i$ with a relatively small number of challenge queries. We find that commercial Iridium devices still rely on this weak algorithm, making them susceptible to SIM cloning attacks. Furthermore, unlike modern cellular systems, the Iridium protocols currently in use neither encrypt traffic by default nor provide mutual authentication, i.e., the satellite does not authenticate itself to the user terminal. This deficiency enables impersonation attacks in which a rogue transmitter can masquerade as an Iridium satellite and deceive user devices.

Our contributions in this work are summarized as follows:
\begin{itemize}
\item \textbf{SIM cloning:} To our knowledge we are the first to explicitly confirm Iridium’s ongoing use of the GSM COMP128-1 hash algorithm for authentication, including its ``butterfly'' structure and known collisions. By using a COMP128-1 cracker and successfully extracting the $K_i$ from Iridium SIM cards, we demonstrate full SIM cloning and authentication bypass in the Iridium network. We also validate the cloned SIM card by using it for an authentication process in the network.
%We also analyze why the COMP128-based authentication is vulnerable to %brute-force and collision attacks, and how an attacker can exploit these flaws over the Iridium radio link.

\item \textbf{Frame Interception and Reassembly:} We develop a pipeline to capture Iridium L-band traffic with software-defined radios and to process the raw Iridium bursts into higher-layer messages. Building on open-source tools \texttt{gr-iridium} for demodulation and the \texttt{iridium-toolkit} parser, we introduce additional methods to associate fragments of the same communication session (e.g., voice call or IP data session) across time and frequency. This allows us to reassemble complete user messages and data streams from the disjoint bursts that the Iridium TDMA/FDMA network transmits.

\item \textbf{Analysis of Iridium Traffic:} Using a distributed set of ground receivers (the LeoCommon observatory network~\cite{leocommon}), we analyze over a month of Iridium downlink traffic covering millions of frames. Our analysis reveals that 88.5 \% of the captured frames on the Iridium radio link have low entropy and thus are unencrypted. These include protocol signalling frames as well as user services (voice, messaging, data). Among these, we identified interactive protocols that do not encrypt login credentials and coordinates, enabling potential account takeover or location tracking by eavesdroppers.

\item \textbf{Spoofing and Jamming Attacks:} We built a GNU Radio-based Iridium signal encoder (\texttt{gr-iridiumtx}) to generate legitimate-looking Iridium signals for attack experimentation. We demonstrate a \emph{downlink spoofing} attack in which a nearby SDR transmitter impersonates an Iridium satellite and sends false control messages to a target device. Because the Iridium handset does not authenticate the network, it accepts messages from the spoofed transmitter (e.g., false pager messages or ring alerts). We also evaluated \emph{jamming attacks}: due to Iridium’s low received signal strength on Earth, even a handheld transmitter can overpower the genuine signal in a local area. We show that by transmitting a continuous waveform or valid-looking dummy frames to jam Iridum Ring Alerts on Iridium’s downlink frequencies, an attacker can effectively disrupt Iridium service (e.g., preventing calls or messages) within line of sight. Our measurements align with recent theoretical studies that quantified the jamming power needed to suppress Iridium signals.
\end{itemize}

Taken together, our findings paint a dark picture: Iridium’s current radio link %(excluding ``Iridium NEXT'' enhancements) 
offers little resistance to attackers. Passive adversaries can intercept and decode global communications, while active attackers can impersonate network entities or deny service entirely. 

In the remainder of this paper, we elaborate on these results and put them in context. Section 2 provides background on the Iridium system architecture and its GSM-derived protocols. Section 3 describes the COMP128-1 authentication algorithm and our reverse-engineering of Iridium’s authentication mechanism. Section 4 defines our attacker model and assumptions. Section 5 outlines the experimental setup for key extraction, traffic capture, and signal injection, respectively. In Section 6 we present the results of SIM cloning, traffic analysis, and attack experiments. Section 7 discusses the broader implications for satellite security and lessons for the design of a more secure Iridium. Section 8 surveys related work on satellite and GSM security. Finally, Section 9 concludes.

\section{Background}

This section presents our analysis of the internal architecture and protocols of Iridium based on information from leaked technical reports~\cite{iridium-security-guide}, developer guides, system requirements specifications, API documentations, presentations by researchers at hacker conferences~\cite{ccc-iridium}, and findings from our own reverse engineering efforts.

\subsection{Iridium Network Architecture}
Iridium is a LEO satellite phone network with global coverage. The first-generation Iridium constellation, deployed in 1997–1998, consists of 66 active satellites in near-polar orbits at an altitude of about 780 km. The satellites are arranged in 6 orbital planes with 11 satellites each, ensuring that at least one satellite is always overhead anywhere on Earth. Each Iridium satellite projects 48 spot beams on the Earth’s surface, dividing its footprint into cell-like regions about 600 km in diameter. The spot beams have a fixed orientation on the satellite, which allows frequency reuse in neighbouring beams. Automatic handover procedures allow users to transfer from beam to beam \cite{jedermann-record-2024}. 

\begin{figure}[t]
\centering
\includegraphics[width=1\columnwidth]{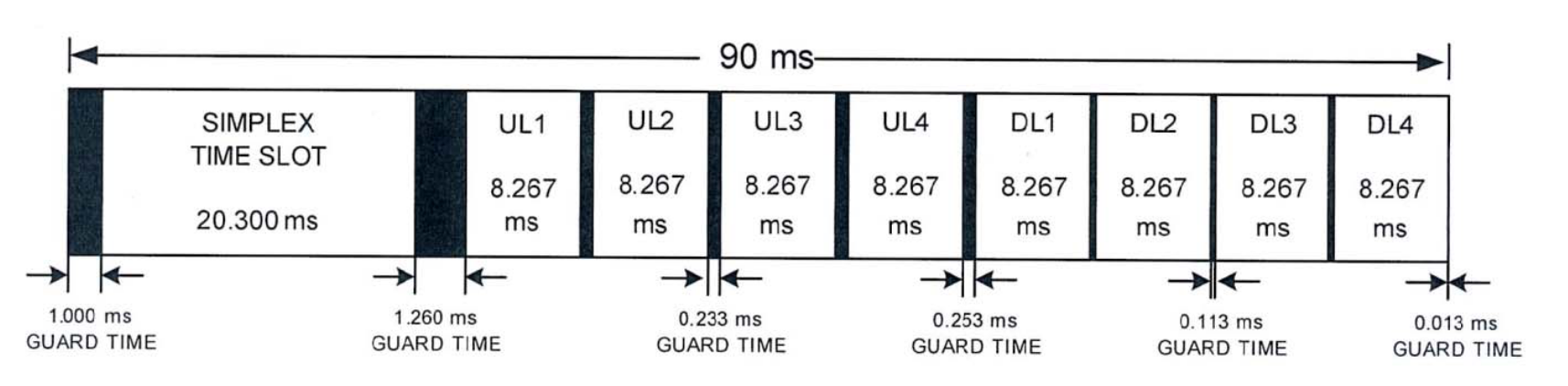}
\caption{Iridium TDMA frame structure \cite{iridium-next-engineering-statement}.}
\label{fig:tdma}
\end{figure}

Iridium satellites communicate with user devices (handsets, maritime terminals, IoT modems, etc.) over L-band frequencies around 1.6 GHz. The user radio link operates in a combined FDMA/TDMA scheme: specifically, 10.5 MHz of spectrum from 1616 MHz to 1626.5 MHz is divided into many narrow-band carrier channels (FDMA), each of which is time-shared among users in TDMA frames (see Figure ~\ref{fig:tdma}). On the legacy Iridium system, each TDMA frame is 90 ms long and contains 4 uplink and 4 downlink time slots, user traffic (phone calls or data) modulates these slots at roughly 25 kbps QPSK, providing a raw data rate of 2.4 kbps per channel after error coding and overhead. The network supports full-duplex by assigning separate time slots for uplink and downlink. In total, the system can support 80 simultaneous voice circuits per spot beam and 172,000 simultaneous users spread across the constellation. User devices access the network by first acquiring a control channel (used for initial registration, incoming call alerts, etc.), then are assigned traffic channels for active calls or data sessions~\cite{pratt2009operational}.

\subsection{Device Authentication}

\begin{figure}
    \centering
    \includegraphics[width=1\columnwidth]{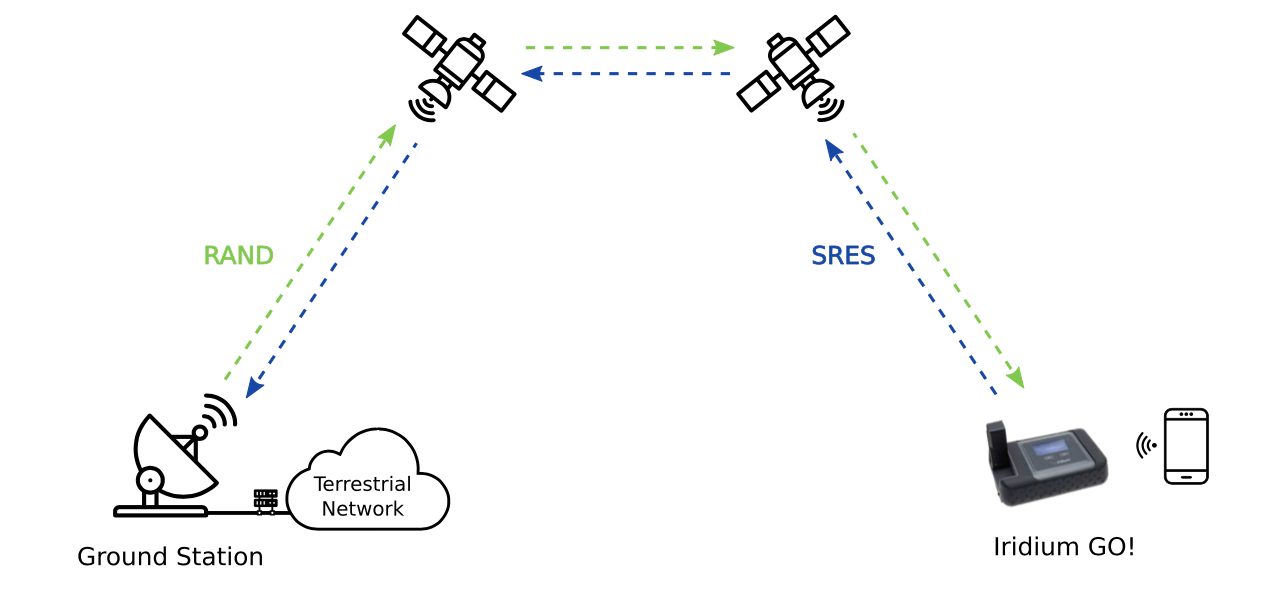}
    \caption{Authentication process in the Iridium network.}
    \label{fig:auth-iridium}
\end{figure}

Iridium user devices contain a removable SIM card (also called an Iridium Access Card) very similar to a GSM SIM, which stores the subscriber’s unique identifier and secret key for authentication. To connect to the Iridium network, a device must execute a GSM-style challenge-response authentication using this SIM. If authentication succeeds, the device is ``registered'' and can send or receive calls/data through the satellite. Importantly, while Iridium’s protocol was modeled after GSM, it did not adopt GSM’s A5 voice encryption on the air interface. 
Instead it relies on the COMP128 family of algorithms that implement the GSM A3 and A8 functions: given the secret $K_i$ and a random challenge, it outputs a signed response $SRES$ (A3) and a session key $K_c$ (A8). The original COMP128-1 algorithm, used in early GSM and, apparently, in Iridium, is a hash function with a 256-bit input (128-bit $K_i$ concatenated with 128-bit $RAND$) and a 128-bit output. The output’s most significant 32 bits are the SRES, and the last 64 bits (with 10 static zero bits at the end) form the $K_c$~\cite{Veeneman2025decode}. 

%\autoref{fig:comp128-io} depicts this input/output structure. 

COMP128-1 was designed to be efficient for SIM cards, but its design was kept secret until 1998 when Briceno et~al. \cite{briceno1998gsm} cracked and published it. An intentional weakness, is that $K_c$ is effectively only 54 bits, making brute-force attacks on the cipher easier.

\begin{figure}[t]
\centering
\includegraphics[width=0.88\columnwidth]{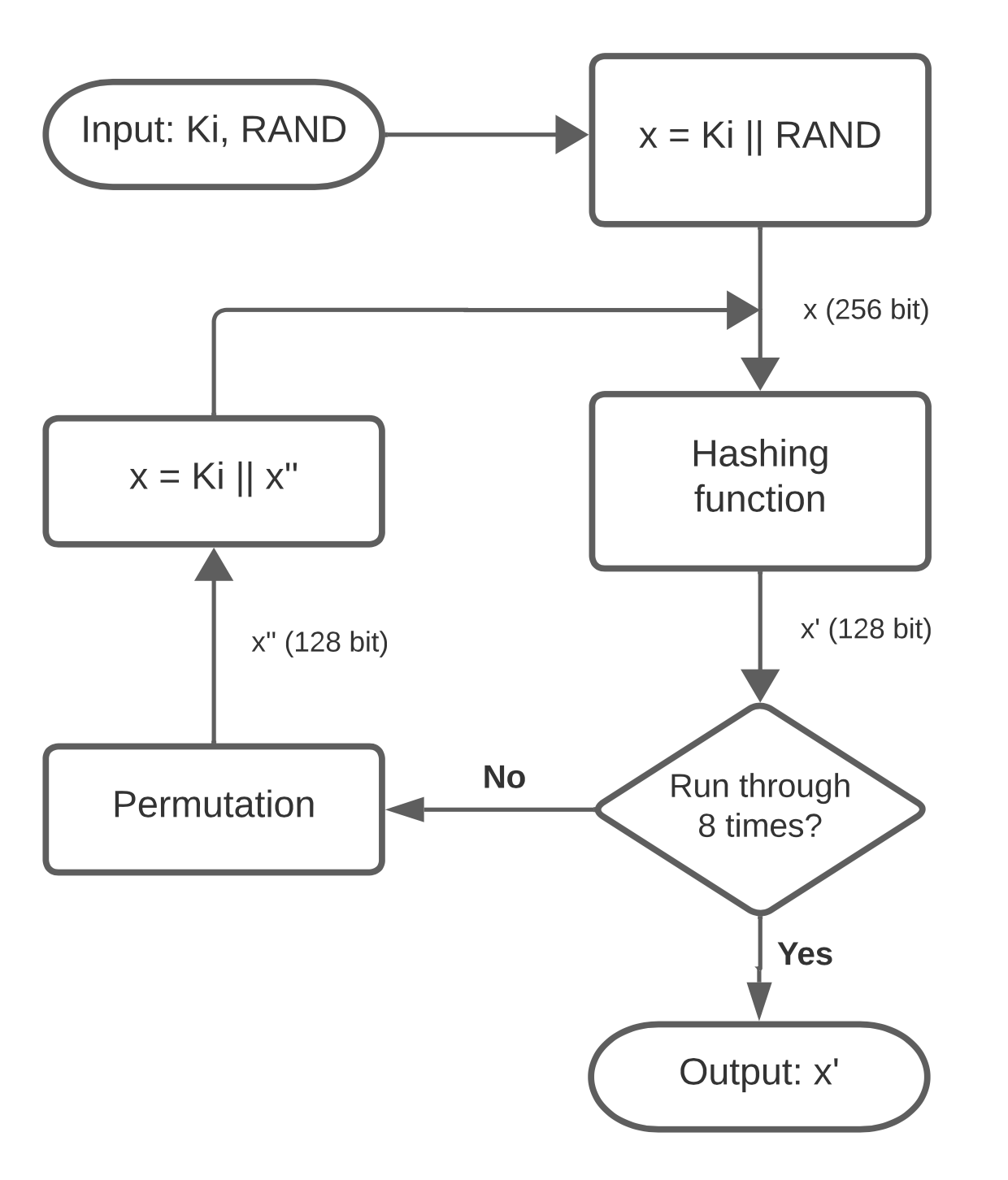}
\caption{Flowchart of COMP128-1.}
\label{fig:comp128-flow}
\end{figure}

Figure \ref{fig:comp128-flow} shows a flowchart of the COMP128-1 hashing process, with eight rounds iterating a compression-permutation sequence.
The internal hashing function uses a butterfly structure in each layer, mixing 32 values of 8-bit. Between each of the 5 layers, the values are combined in pairs (using modulo arithmetic and table lookups), while each table lookup shortens the values by 1 bit, ending with 32 4-bit values. The algorithm is surjective but not injective, i.e., many inputs can produce the same output (especially due to the zeroing of the last 10 output bits). This non-injective property leads to collisions that have been exploited in attacks \cite{briceno1998gsm}. 

%\begin{figure}[t]
%\centering
%\includegraphics[width=0.98\columnwidth]{comp128_flowchart.png}
%\label{fig:comp128-flow}
%\end{figure}

The COMP128-1 algorithm is vulnerable to chosen-challenge attacks and collision attacks that recover $K_i$ with about 150,000 challenges. \cite{briceno1998gsm} Shortly after its public revelation, researchers started optimizing this attack by prioritizing carefully chosen RAND values that are more effective in causing collisions. Also replacing challenges for the last digits by offline brute force calculations improved the efficiency since the SIM card is the bottleneck. With those optimizations, $K_i$ can be cracked with as few as 20,000 queries, which should take about one hour \cite{wray2003comp128, olawski2011security}. %Even a non-adaptive brute force on COMP128-1 is feasible since the effective key space is $2^{54}$ due to the $K_c$ zeroing. 
Our experiments confirm the $K_i$ extraction and SIM cloning of GSM. With new SIM card readers our $K_i$ extraction was significantly faster than the 1h prediction. Given that Iridium SIMs process unlimited authentication requests (i.e., no counter to throttle repeated RAND challenges), an attacker with temporary physical access to an Iridium SIM could perform the attack to extract $K_i$.

\subsection{Protocols}

\begin{figure}[ht]
    \centering
    \includegraphics[width=1\columnwidth]{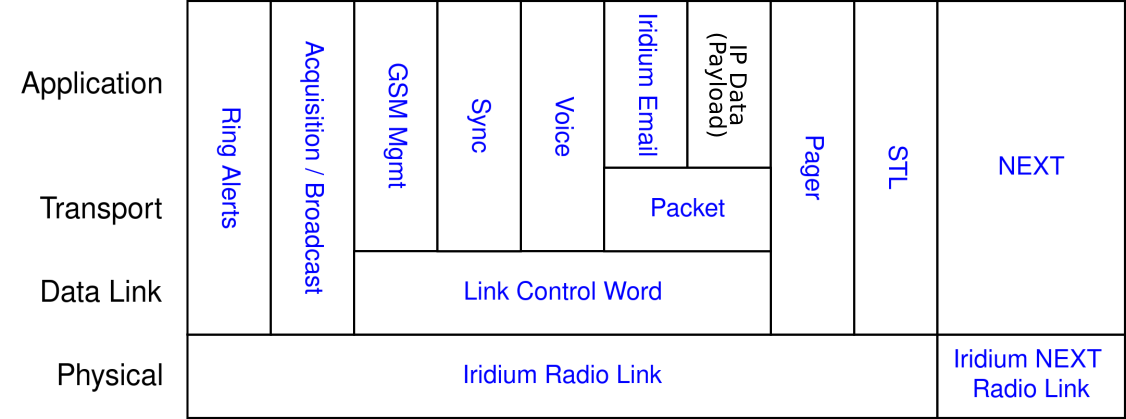}
    \caption{Iridium downlink protocol stack}
    \label{fig:protocol-stack}
\end{figure}

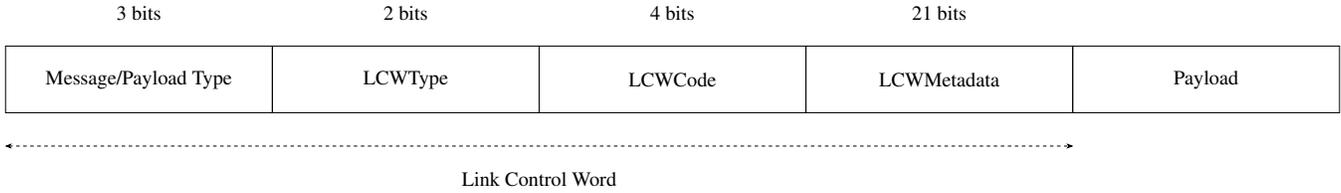
\begin{figure*}[!ht]
\centering
\resizebox{1\textwidth}{!}{%
\begin{circuitikz}
\tikzstyle{every node}=[font=\LARGE]
\draw  (0,16) rectangle (8,14);
\draw  (8,16) rectangle (16,14);
\draw  (16,16) rectangle (24,14);
\draw  (24,16) rectangle (32,14);
\draw  (32,16) rectangle (40,14);

\draw [<->, >=Stealth, dashed] (0,13) -- (32,13);
\node [font=\LARGE] at (4,15) {Message/Payload Type};
\node [font=\LARGE] at (4, 17) {3 bits};
\node [font=\LARGE] at (12,15) {LCWType};
\node [font=\LARGE] at (12,17) {2 bits};
\node [font=\LARGE] at (20,15) {LCWCode};
\node [font=\LARGE] at (20,17) {4 bits};
\node [font=\LARGE] at (28,15) {LCWMetadata};
\node [font=\LARGE] at (28,17) {21 bits};
\node [font=\LARGE] at (36,15) {Payload};
\node [font=\LARGE] at (36,17) {};
\node [font=\LARGE] at (16, 12) {Link Control Word};
\end{circuitikz}
}%
\caption{Link Control Word (LCW) format}
\label{fig:lcw-header}
\end{figure*}

The Iridium network has many protocols and services that have been developed over the years since the emergence of the constellation. The new Iridium NEXT satellites (completed in 2019) maintained backward compatibility with legacy protocols while upgrading satellite hardware and adding new services. 
The Iridium protocol stack can be seen in Figure \ref{fig:protocol-stack}. The blue names indicate proprietary protocols that are specific to Iridium. 
%The original Iridium Radio Link is defined by the TDMA structure shown in \ref{fig:tdma}. The burst duration for the uplink and downlink slots is 8.267 ms. The burst consists of a preamble that spans 16 symbols, a unique word, or a synchronization word, which are 12 symbols that are separately unique for uplink and downlink. The payload is then appended, and both the unique word and the payload are encoded using differential QPSK. 
The legacy Iridium and the Iridium NEXT Radio Link have the same TDMA structure; however, Iridium NEXT allows for more wideband transmission spanning multiple subcarriers, which results in higher symbols per second. Thus, the channels vary from 41 to 200 kHz. %In addition, the preamble is included in the unique word. The modulation of the burst can be differential QPSK or 16-APSK \cite{certus-bcx-subsystem-req}.
In both versions, the satellite network that handles the routing encapsulates the traffic, so it is not visible for user terminals or eavesdroppers.
%In both versions, the data link layer is represented by the Link Control Word (LCW), which is a header/payload structure encoded in the Radio Link payload. 

When accessing the network, a terminal first uses the Ring Alerts (on the left of Figure~\ref{fig:protocol-stack}) to find the beam-dependent broadcast channels. Those are used for synchronization and exchanging acquisition messages. After this, a data link between the satellite and the terminal is established. The link control protocol is used to manage the connection. Different Link Control Words (LCWs) as shown in Figure~\ref{fig:lcw-header} implement different types of channel management tasks, such as authenticating the terminal with GSM management messages. Other LCWs are monitoring the physical signal characteristics, manage handovers, ensure synchronization, and transmit voice call data. Above the data link, an Iridium-proprietary packet protocol similar to TCP is available, offering reliable data transfer by implementing sequence numbers and acknowledgements. On top of them, common application layer protocols such as mail or HTTP can run to enable classical Internet services over Iridium radio links. 

\section{Attacker Model}
In our threat model, the adversary is equipped with SDRs and computing resources that are readily available to hobbyists. We assume the attacker’s goal is to compromise Iridium communications in one or more of the following ways: (1) eavesdrop on confidential communications (voice calls, messages, data) of target users, (2) impersonate a legitimate Iridium device or network (spoofing) to inject false messages or gain unauthorized access, (3) disrupt Iridium service availability for users (jamming or denial-of-service), and (4) track the user location (privacy). Below we detail the attacker’s capabilities and the specific assumptions.\\

\textbf{Attacker Capabilities:} The attacker has one or more radio transceivers capable of tuning to the Iridium frequencies (1616–1626.5 MHz) in the L-band. For example, an inexpensive SDR like the HackRF One, coupled with a suitable L-band antenna, can transmit and receive Iridium radio signals. The attacker can thus passively intercept traffic from satellites or users on Earth, as well as transmit arbitrary data frames to those entities.\\ %Importantly, Iridium’s downlink signals are relatively strong (designed to reach handhelds outdoors), and one satellite’s beam covers a large footprint (up to ~1000 km diameter). This means a single eavesdropper could potentially monitor traffic of many users across a region. Uplink signals (from users to satellite) are much weaker and can only be received if the attacker is fairly close (within tens of kilometers) to the transmitting device, due to the device’s low transmit power and the satellite antenna focusing the uplink towards the sky. Our model assumes the attacker primarily intercepts downlink traffic, which carries both directions of communications (since Iridium uses bent-pipe satellites that immediately downlink what they receive). Uplink eavesdropping is considered opportunistic (e.g., if the target is nearby or using a high-gain antenna terminal).

%We also assume the attacker can \textbf{transmit} arbitrary signals in the Iridium band. This requires an SDR or signal generator capable of L-band transmission and potentially an amplifier. With a directional antenna or proximity to the victim, even a low power device may overpower the satellite’s signals.\\ %The attacker could thus conduct jamming by outputting noise or continuous signals. For spoofing, the attacker can generate signals that mimic Iridium’s TDMA bursts in structure (e.g., correct burst timing and modulation). The open-source \texttt{gr-iridiumtx} we use was developed for exactly this purpose: to encode raw bits into Iridium GMSK-modulated bursts at the right symbol rate and burst shape.

%We assume the attacker does \emph{not} have physical access to the satellites or gateway infrastructure, and does not possess any internal Iridium encryption keys or proprietary software (aside from what can be extracted via public means). Essentially, the adversary is operating externally via the radio interface, which aligns with realistic scenarios (e.g., pirates, hackers, or espionage agents on the ground).

\textbf{Attacker Knowledge:} The attacker may know the target device’s Iridium phone number or IMSI (if targeting a specific user). They may also know approximate location or timing of the target’s communications (for instance, if trying to spy on a known ship at sea using Iridium phones). Some Iridium attacks (like traffic interception) do not require knowing the target’s identity in advance, as the attacker could simply collect and examine all traffic for interesting content. %Other attacks, like impersonation, benefit from having extracted a specific user’s credentials or $K_i$ in advance.

We assume the attacker has access to general technical knowledge of Iridium’s protocols. Specifically, they know how to decode the frame structure, how to parse signaling messages, and how to encode messages for transmission. They may also have reference implementations or open libraries for Iridium encoding/decoding (indeed we leverage the \texttt{iridium-toolkit} and related software published open source by Schneider and Zehl and the Chaos Computer Club~\cite{schneider2022gr}).\\

\textbf{Scope:} Our threat model does not consider attacks on the gateway or control channels beyond the Radio Link layer (e.g., hacking into the Iridium ground station or injecting data into the terrestrial network). We focus on attacks executed over the wireless radio link. We also do not consider higher-layer cyber attacks on user terminals that exploit software vulnerabilities (e.g., malware on an Iridium GO! device), which have been discussed in~\cite{santamarta1}.\\

%In summary, the adversary in our model is an RF attacker with moderate resources who can intercept and transmit Iridium L-band signals. This is a plausible adversary given the proliferation of SDRs and known interest of criminal or state actors in satcom interception. Recent work has shown even hobbyists can intercept satellite phone calls and pager messages due to lack of encryption. The unique aspect for Iridium is achieving \emph{real-time injection} and targeted disruption, which we address by custom transmitter development.

\textbf{Attack Primitives:} 
Given these capabilities, the attacker’s goals translate to the following concrete attack scenarios, which we will investigate in this paper:\\

    \emph{Eavesdropping attack:} silently collect downlink or uplink traffic and decode it to retrieve possibly sensitive information (violating confidentiality).\\

    \emph{SIM cloning attack:} extract $K_i$ from target’s SIM (via physical means) and use it to impersonate the target on the network (violating authentication).\\

    \emph{Spoofing attack:} set up a fake Iridium satellite station that sends control messages to a target device, perhaps to manipulate or phish the user (violating integrity/authenticity).\\

    \emph{Replay attack:} record an authentication exchange from a legitimate session and replay it to gain network access for a clone device, or to desynchronize the target.\\

    \emph{Jamming attack:} transmit noise or crafted signals to deny Iridium service in an area (violating availability). Specifically, block incoming ring alerts or disrupt ongoing sessions.\\

    \emph{Location tracking:} Determine the approximate location or movement of a specific Iridium subscriber (violating privacy). \\
    %Prior research shows a) that ring alert messages contain a satellite and beam ID that can reveal the user’s location down to an small radius with enough observations \cite{jedermann-record-2024,liu2025mind} and b) that Iridium devices provide some location data on logging on the network \cite{ccc-iridium}.

\section{Experimental Design}
In this section, we detail the tools and setups used to evaluate the security of Iridium. Our experiments comprised three main parts: (1) authentication key extraction (for SIM cloning), (2) traffic interception and analysis, and (3) active signal injection (for spoofing and jamming).

\begin{figure}[ht]
    \centering
    \begin{subfigure}{.26\linewidth}
\centering
\includegraphics[width=1\linewidth]{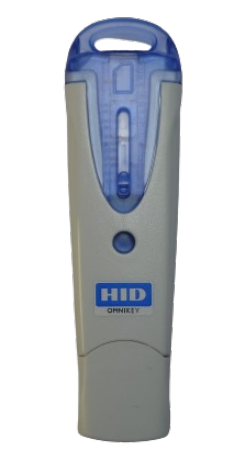}
\subcaption{OMNIKEY}
\label{fig:card_readers:omnikey}
\end{subfigure}% 
\hfill
\begin{subfigure}{.48\linewidth}
\centering
\includegraphics[width=1\linewidth]{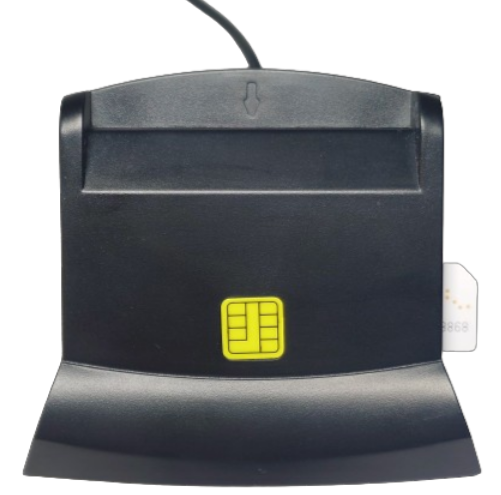}
\subcaption{CSL}
\label{fig:card_readers:csl}
\end{subfigure}% 
    \caption{The used SIM card readers}
    \label{fig:card_readers}
\end{figure}

\subsection{Key Extraction and SIM Cloning Setup}
To extract the COMP128-1 secret key ($K_i$) from an Iridium SIM card, we tested two standard SIM card reader \textit{HID OMNIKEY 6121 Mobile}, shown in Figure~\ref{fig:card_readers:omnikey}, and \textit{CSL - USB Smart Card Reader}, in Figure~\ref{fig:card_readers:csl}. For extracting the key $K_i$ we used the program `Woron Scan v1.09', which is running on a laptop with windows 10. %We wrote a program that sends \texttt{GSM\_Authenticate} APDU commands to the SIM, which trigger the COMP128-1 computation on the card. Since Iridium SIMs use the same command set as GSM, we follow the GSM protocol semantics. 

The program has a set of prepared RAND challenges designed to exploit the birthday paradox collision technique following the method proposed in~\cite{briceno1998gsm}). Essentially, the used pairs of RANDs often differ only in a few bits, prioritizing those with a high collision probability. The goal was to create collisions in the first stages of the butterfly network inside the hashing function of COMP128-1. By comparing the SRES outputs for those challenge pairs, one can deduce certain bits of $K_i$. We verified the extracted $K_i$ by comparing the SRES outputs from the original SIM card and from an open source COMP128-1 implementation~\cite{Munaut2009comp128v1}. For this we feeded the open source implementation with the previously extracted $K_i$ and a couple of random challenges. The same challenges were given to the SIM card and both results were compared.

For SIM cloning, we used a blank programmable SIM (specifically a \textit{SIM MAX 12 in 1 Card} that can be configured with up to 12 arbitrary IMSI and $K_i$). We programmed the IMSI and the extracted $K_i$ into the blank card byusing `SIM Scanner v5.15'. For verification of the new cloned SIM we did two independent tests: First, we used one of our card readers to send a couple of random challenges to the original SIM, than we swapped the card with the cloned SIM and send the same challenges. In the second test, we inserted the clone into an Iridium GO! handset and attempted to register on the network. The handset successfully registered, indicating the network accepted the cloned credentials. To avoid conflicts, we made sure the original SIM/device was powered off during this test. This experimental setup confirmed that an extracted $K_i$ is all that is needed to impersonate the subscriber—Iridium does not employ any additional challenge beyond GSM’s.

\begin{figure*}
    \centering
    \includegraphics[width=1\textwidth]{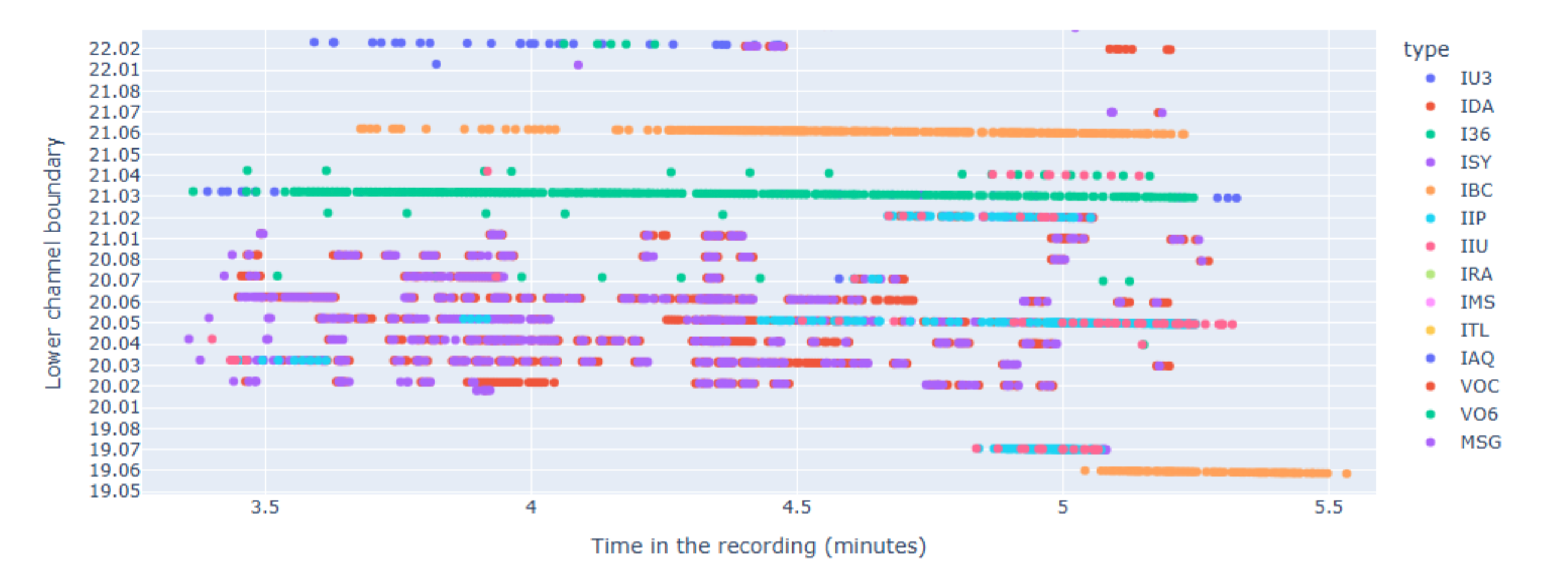}
    \caption{Detail showing “lanes” and FDMA channel drift. The Y-axis is plotted by frequency in Hz
but labelled as FDMA channel access}
    \label{fig:iridium-fdma}
\end{figure*}

\subsection{Traffic Capture and Processing}
For passive traffic capture, we deployed two different Iridium receivers based on low-cost SDRs. Our first receiver used a HackRF One with an L-band passive wall-mounted antenna (Taoglas)~\cite{taoglas2021external} on an Intel NUC. Our second receiver consisted of an Ettus USRP B210, with an active HC610 antenna from Callian~\cite{calian2023HC610}. This antenna is supplied with 12V DC power through a bias tee. The setups were located in an open area with clear sky view. Additionally, we received limited access to the crowdsourced LeoCommon network \cite{leocommon}, which supports capturing Iridium traffic from up to 10 ground stations.

To demodulate the Iridium signals, we ran the \texttt{gr-iridium}~\cite{schneider2022gr} GNU Radio out-of-tree module. This module outputs a stream of ``.bits' files, each containing raw frames with metadata (timestamp, frequency, confidence). To decode  the captured frames, we used the open-source \texttt{iridium-toolkit} that parses Iridium bursts into structured messages. This toolkit recognizes frame types like Ring Alert (IRA), voice traffic, SBD (Short Burst Data), and so on. 

For voice calls, frames are continuous and contain vocoder bits (Iridium uses a proprietary voice codec called AMBE). Decoding actual audio was beyond our scope, but we note it has been done on real-world data by hackers and hobbyists since Iridium voice calls are  not encrypted \cite{ccc-iridium}. For data sessions (e.g., Iridium IP or SBD bursts), we concatenated the payloads from sequential frames after removing any filler or CRC bits as specified by the protocol. In many cases, the result was a packet containing HTTP or an AT-command style message in plain ASCII. For instance, we could reconstruct our own mobile-originated SBD messages containing text or GPS coordinates, and in some cases IP data sessions that were part of an Internet connection (e.g., an email over dial-up PPP). 

From these initial observations of the data we could encounter plaintext traffic of users, thus we decided to develop a privacy-preserving pipeline, that would analyse the metadata of the traffic, without preserving any data present in the observed Iridium traffic. To achieve this we do all of the processing in memory using a \texttt{gr-iridium} process, which buffered output is piped to a instances of parser from \texttt{iridium-toolkit} that decodes and categorizes the bitstreams into simple text frames or LCW frames. From these frames the pipeline extracts the information such as frame type, channel assignment. At this stage the single frames are dropped and the LCW frames are reconstructed into longer messages. 
Our approach was as follows: since Iridium downlink frames include the frequency channel and TDMA slot in their metadata, frames from one user’s session tend to appear on the same frequency (or a small set of frequencies) and at regular TDMA slot intervals (every 90 ms frame). We clustered frames by their frequency and timing pattern. Figure \ref{fig:iridium-fdma} illustrates how frames form distinct ``lanes'' in a time-frequency plot corresponding to different calls or data sessions, and how clustering can isolate them. As a result we can partially reconstruct the messages and calculate their entropy. At the end of this we drop the buffered frames and reconstructed frames.

% ********************************************************************************

To determine if a data session is encrypted or not, we rely on the Shannon entropy.  
While computing the Shannon entropy over the different data sessions, we found a bimodal distribution: data sessions with high entropy ($>7$ bits/byte), indicative of either compressed or encrypted data, while sessions with low entropy in the $4$–$5$ bits/byte range, strongly suggesting text or other types of non-encrypted

\subsection{Radio Frequency Attack Testbed}\label{sec:spoof-setup}
\begin{figure}
    \centering
    \includegraphics[width=1\linewidth]{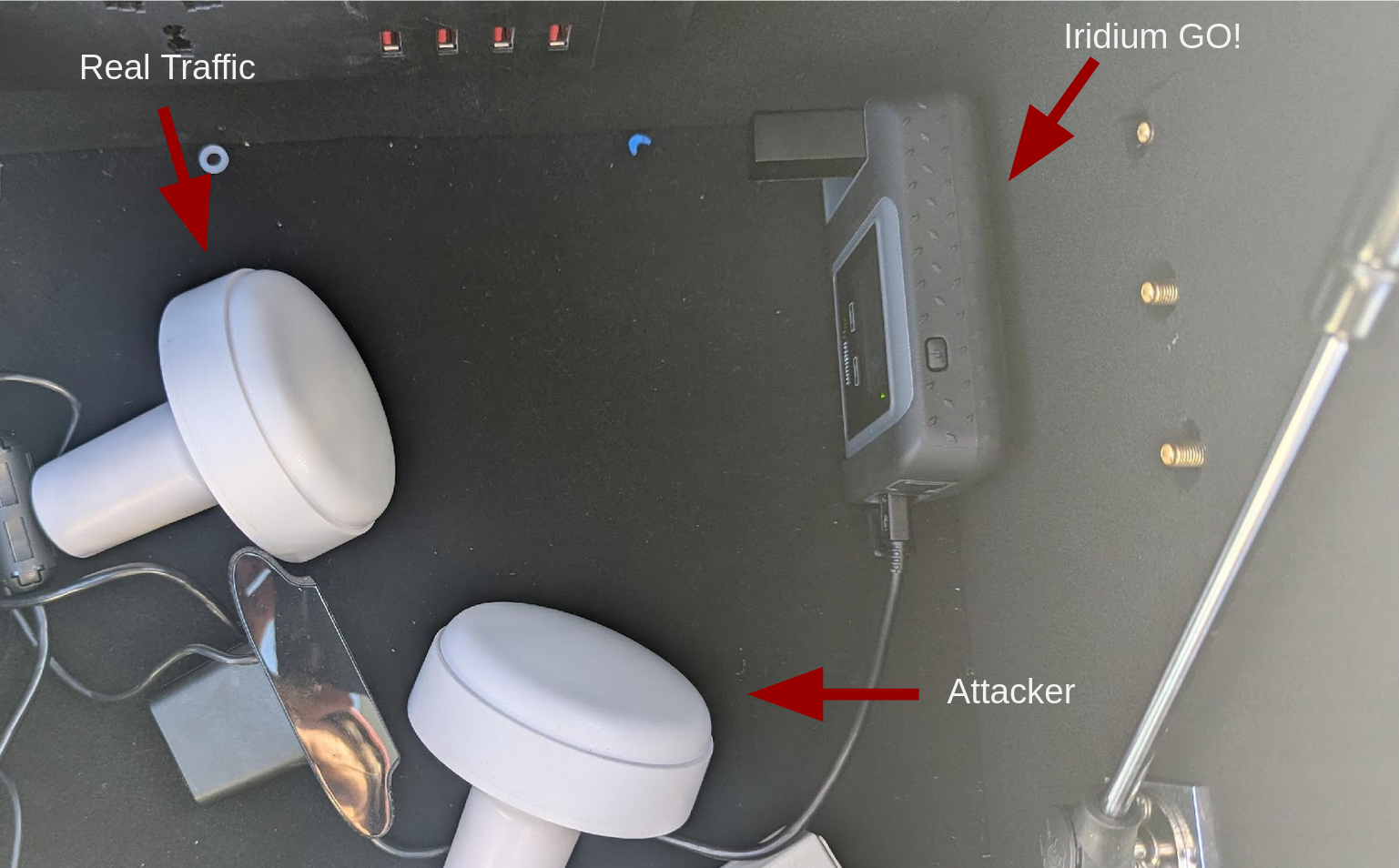}
    \caption{Attacker setup inside Faraday cage.}
    \label{fig:attacker-setup}
\end{figure}

For our active attack experiments, we built a controlled test environment to avoid interfering with real users. Throughout these tests, we took care to operate in a shielded anechoic chamber or in a Faraday cage to avoid unintended leakage. We also had spectrum analyzers monitoring to confirm we weren’t transmitting beyond the necessary duration or band. We used an Iridium device,
called Iridium GO!, that supports SBD, SOS, SMS, web data, and Voice calling Iridium services \cite{iridium-go}. It uses the Iridium Core 9523 as the Iridium module. We used
Iridium GO! App on an android smartphone to connect to Iridium GO! via a Wi-Fi connection and use all the available functionality of Iridium GO!. Figure \ref{fig:attacker-setup} illustrates our experimental setup inside the Faraday cage.

Our transmit setup consisted of an SDR (HackRF One) capable of transmitting in L-band, an active Iridium antenna with a bias-tee powered with 12V and a pair of directional patch antennas pointed at the target device.

We crafted several types of signal injections, which we implemented as a new tool called \texttt{gr-iridiumtx} to realize the following tasks.\\

\textbf{Replay bursts:} By recording the authentication of the Iridium GO! against the Iridium network, we perform radio signal replay attacks that retransmit the authentication sequence. To provide recordings of both the uplink and the downlink at the same time, the user device cannot be placed too close to the recording device, otherwise, the uplink will overpower the downlink signal, and thus the attacker recording device was placed at a distance of 560 m from the Iridium GO!. \\

\textbf{Synthetic ring alert:} We generate completely synthetic Ring Alert messages and send them to our test device. We encoded the bursts in the exact same format as a normal Ring Alert frame. \\

% \textbf{Downlink spoofing:} 
    %move to results (if at all)
    %In our tests, the phone did not fully process our fake assignment (likely because we did not perfectly mimic all the framing of the sync channel), but it clearly received and attempted to respond before failing. This partial success shows that with more refinement, a man-in-the-middle base station could coerce a device to reveal authentication responses—essentially acting as an IMSI-catcher for Iridium. Max Lill reported a similar replay attack feasibility when network authentication is absent.
    
\textbf{Jamming signals:} 
For the experimental jamming attack, we rely on a continuous Gaussian noise signal sent over 500 kHz spectrum on the Iridium L-band link at 1626.25 MHz center frequency. The attacker is placed at a fixed distance from the Iridium GO! while varying the transmission power of the attacker radio device.

\section{Results}\label{sec:results}
We now detail the results of our security evaluation of Iridium, organized by the type of vulnerability. Table \ref{tab:attacks} provides an overview of the attacks we performed and their outcomes.

\begin{table*}[t]
\centering\small
\begin{tabular}{p{4.5cm}p{5.5cm}p{5.5cm}}
\hline
\textbf{Attack/Vulnerability} & \textbf{Description \& Outcome} & \textbf{Result} \\
\hline
$K_i$ Extraction (COMP128-1) & Retrieve SIM secret by sending challenges to SIM. & \textbf{Success}: Recovered $K_i$ from Iridium SIM in 6 minutes; cloned SIM worked on network. \\\\
%SIM Cloning & Impersonation & Use extracted Ki to clone SIM and register as victim. & \textbf{Success}: Cloned SIM authenticated to Iridium, enabling impersonation (calls, data) as victim. \\
Eavesdropping (Downlink) & Passively capture and decode unencrypted downlink traffic frames. & \textbf{Success}: Captured $>$186 million frames; unencrypted voice, SBD, IP data. \\\\
%Cleartext Sensitive Data & Presence of unencrypted personal or security-critical info in traffic. & %\textbf{Success}: Found credentials, emails, coordinates, and government data in plaintext. Account passwords observed. \\
Uplink Interception & Receive uplink transmissions from user devices. & Partial: Only if near target (within $\sim$20–30 km). Rarely observed in our remote dataset. \\\\
Location Tracking & Determine user’s location from downlink signals (paging beams). & \textbf{Success}: Using ring alert beam IDs over time narrowed location to $\sim$10 km radius. Also tracked movement across beams. \\\\
Spoofing (Downlink Impersonation) & Transmit fake network messages to device (no mutual authentication). & \textbf{Success}: Device accepted fake ring alert. Partial success impersonating authentication sequence (device responded). \\\\
Replay Attack & Record and replay network messages. & \textbf{Success}: Replayed a previously captured authentication sequence; device authenticated itself against the attacker. \\\\
Jamming & Jam channel with minimal power. & \textbf{Success}: Continuously sending dummy bursts on control channel prevented registration and ring alerts. \\ %at J/S -5 dB. \\
\hline
\end{tabular}
\caption{Summary of attacks on Iridium and their outcomes in our experiments.\label{tab:attacks}}
\end{table*}

\subsection{Authentication and SIM Cloning Results}
\begin{figure}
    \centering
    \includegraphics[width=1\linewidth]{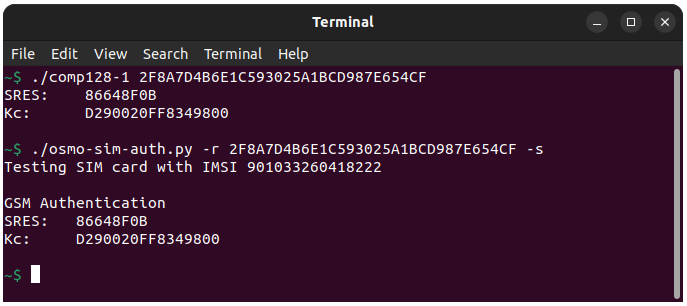}
    \caption{Open source algorithm (1st command) vs. SIM card (2nd command), using the same challenge}
    \label{fig:key_extraction}
\end{figure}

Our tests confirmed that the Iridium authentication scheme is essentially the GSM COMP128-1 challenge-response and is vulnerable to known attacks.
Using the Omnikey card reader and Woron Scan we were able to recover the secret key $K_i$ with 20,711 queries, which was completed in only 6 minutes. The CSL card reader also only took just under 10 minutes. In both cases the program terminated with the recovered 128-bit $K_i$ and IMSI as output. In Figure \ref{fig:key_extraction}, the first command shows the execution and result of the open source COMP128-1 algorithm using the 128-bit challenge 0x2F8...4CF. The key $K_i$ is inserted directly in the code. The second command in Figure~\ref{fig:key_extraction} shows the query on the SIM card, using the same challenge and returning the same SRES (and $K_c$).
%In fact, due to the 10 zero bits in the algorithm’s output, the effective key space is $2^{54}$, so significantly fewer queries (on the order of 8 adaptive queries as per Biryukov et al.) would have sufficed. 
%The extraction required negligible computational effort, 
We also confirmed that Iridium SIMs do not have any defence against rapid-fire authentication attempts, like rate-limiting or a finite attempt counter. Some GSM SIMs later introduced an authentication counter, typically around 60,000, to limit how often the \textit{RUN\_GSM\_ALGO} command can be executed inside a SIM. To ensure the absence of this counter-measurement we stressed the SIM with more than 200,000 queries.

After obtaining $K_i$, we programmed a blank SIM with the same IMSI and $K_i$. The cloned SIM was inserted into an Iridium GO! while the original SIM was off. The clone connected to the network with no issues. %We were able to make a test call from the device with cloned SIM to a regular phone;
We were able to test the cloned SIM by sending and receiving mails via the device and the Iridium network; the procedure was completed normally, meaning the network had authenticated the clone and granted access. This demonstrates a complete break of Iridium’s authentication: an attacker who gets $K_i$  can create their own device to use the victim’s service or impersonate them, using Iridium on the victims cost.

% We also tried to clone a SIM via an over-the-air attack. 
%In a controlled and environment, we set up our SDR to act as a fake satellite. We lured the target device to authenticate with us by impersonating a network broadcast (exploiting the lack of network authentication). When the device sent us an SRES (in response to our challenge), we simply forwarded that to the real network’s challenge (which we had intercepted from a real satellite). This is essentially a classic MITM: we answered the network’s challenge using the device’s computed response. The device was then authenticated to the real network, but through our intermediary. This is a form of replay attack: since Iridium doesn’t bind an authentication to a specific session or have mutual checks, one can replay the authentication messages in a shifted context. %Max Lill observed that user authentication in Iridium is susceptible to replay for similar reasons. 
%In practice, this means an attacker with a brief window can hijack a session or trick a device into giving a valid SRES for use later.

In summary, we achieved a full success in extracting the secret key $K_i$ and in cloning the SIM. The network alone offers no defence once the $K_i$ is known. This is a severe vulnerability, as it compromises the fundamental identity mechanism of the system.

\subsection{Downlink Protocol Analysis}

\begin{table}[h]
    \centering
    \begin{tabular}{l r r}
        \toprule
        \textbf{Category} & \textbf{Count} & [\%] \\
        \midrule
        Acquisition & 47 127 & 0.03\% \\
        Messaging (pager) & 3 039 478 & 1.63\% \\
        Voice & 5 135 995 & 2.75\% \\
        NEXT & 7 184 915 & 3.85\% \\
        Ring Alert & 10 186 711 & 5.45\% \\
        STL & 10 843 844 & 5.81\% \\
        IP Data & 24 223 233 & 12.97\% \\
        Broadcast & 26 817 173 & 14.36\% \\
        SBD/GSM & 29 137 018 & 15.59\% \\
        Sync & 31 935 215 & 17.09\% \\
        Unknown & 38 237 477 & 20.47\% \\
        \midrule
        \textbf{Total} & \textbf{186 788 186} & \textbf{100.00\%} \\
        \bottomrule
    \end{tabular}
    \caption{Observed Iridium frames}
    \label{tab:observed-frames}
\end{table}

\begin{table}
    \centering
    \begin{tabular}{l r r r}
        \toprule
        \textbf{Category} & \textbf{Total} & \textbf{High Entropy} & [\%] \\
        \midrule 
        Pager Msgs& 7 545 & 1 992 & 26.40\% \\
        Voice & 37 515 & 17 284 & 46.07\% \\
        IP Data & 65 875 & 43 664 & 66.28\% \\
        NEXT & 319 625 & 117 670 & 36.82\% \\
        SBD/GSM & 1 144 874 & 769 & 0.07\% \\
        \midrule
         \textbf{Total} &	 \textbf{1 575 434} &	 \textbf{181 379} &	 \textbf{11.51\%} \\
        \bottomrule
    \end{tabular}
    \caption{High entropy packets statistics}
    \label{tab:encrypted-packets}
\end{table}

In our efforts to better understand the risks of an attacker passively capturing downlink traffic to decode and analyze its contents, we compute usage statistics of the different Iridium protocols and discuss their likelihood of leaking sensitive or personal identifiable information over these protocols.    

Table~\ref{tab:observed-frames} shows the total number of received frames per Iridium message category that we detected with our experimental setup during a continuous observation period of more than a month between January and March 2025. 
Our detailed protocol analysis revealed that from the top ten Iridium frame types shown in the Table, none of them uses encryption by default, revealing a systematic lack of confidentiality in Iridium radio link protocols.

To better understand the encryption practice of the higher-level protocols, we additionally compute the Shanon entropy of the frames that carry user data in their payloads including messaging, SBD/GSM, IP data, voice, and NEXT. Table \ref{tab:encrypted-packets} shows the statistics of the total and the relative number of packets with high entropy (>7 bits / byte). 
Of around 1.5 million downlink frames analyzed, 88.5 \% had low Shanon entropy, basically suggesting that very little encryption is applied at the higher-level protocols as well. The main exceptions were a subset of traffic from special services: e.g., we observed some encrypted mission data in a few ``DoD'' flagged frames (likely from secure government-only devices which apply their own encryption at higher layers, or the push-to-talk AES service). However, ordinary phone call frames, pager messages, short burst data, and internet IP data sessions are very often in the clear.

Iridium enforces apparently no default over-the-air encryption for messaging, voice or data. The optional voice encryption noted in some documents is likely referring to an old proprietary scrambler that is not enabled by default. Similarly, SMTP traffic sent via Iridium dial-up is sent in cleartext. %we reconstructed text of emails, including addresses and content. Figure~\ref{fig:clear-email} shows an example segment of an email (with personal info redacted) that we reassembled from frames; it was a weather report sent by a maritime user via an Iridium data link. The email content and headers are fully visible.
Perhaps the most critical comes from data sessions. We found instances of protocols that are known to transmit login credentials in plaintext. One pattern involved maritime or remote terminal protocols that establish Point-to-Point Protocol (PPP) sessions over Iridium dial-up to access a satellite data service. In such sessions, PPP Password Authentication Protocol (PPP PAP) authentication packets contain a username and password in the clear. These credentials likely allow access to an Iridium data gateway or email relay. An eavesdropper could thus steal these credentials by passively listening to the downlink protocols and potentially gain further access (like retrieving the user’s messages from the service provider’s server, etc.). The same issue relates to other unencrypted protocols actively in use such as FTP. An observer could passively monitor the username and password to later exploit those credentials to login on the server.

%THis tercepted what appears to be an FTP login to a server (it was part of a SCADA system using Iridium; the FTP username ``adm'' and a 4-digit password were seen).

A particularly concerning issue is the transmission of Iridium GO! device’s web dashboard credentials over HTTP. The Iridium GO! is a hotspot device that uses Iridium to relay data and has a web interface for users. In one recorded session, we saw an HTTP Basic Auth header (Base64 encoded credentials) being sent from our GO! device to some server, presumably carrying the admin login for the device or service. %The credentials decoded to a recognizable default username/password combination. 
This suggests not only are user communications exposed, but even device management commands and credentials leak.
One interesting find: we observed some secure phone protocols that turned out to be using an external encryption device. Specifically, a few voice call streams had very high entropy—they were likely encrypted voice using a device like an add-on scrambler. This shows that some users, aware of Iridium’s lack of encryption, add their own. However, these are apparently rare. Most voice calls have such a low entropy that they must be plain.

%All told, our analysis of the dataset showed that at least \textbf{~14\% of all captured traffic frames contained easily identifiable plaintext ASCII strings}. The rest were largely voice frames (vocoded binary, but still decodable) or binary file transfers (e.g., images or firmware updates being sent via Iridium—these had high entropy but not due to encryption, rather due to being compressed or binary format; we confirmed some by file signatures). Only a very small fraction ($<$1\%) seemed to be actually encrypted content (high entropy and likely not a known file format). Those could correspond to users who employed their own application-layer encryption (e.g., using a VPN over Iridium data, or using an encrypted chat app). 

In conclusion, absent user-added encryption, Iridium offers no confidentiality: any adversary with a moderate antenna and SDR can listen to traffic in a radius of hundreds of kilometers from their position (since a satellite downlink covers a huge area).
To put this in context, this is akin to the pre-1990s analog cellular or early GSM A5/0 (no encryption) situation, but in 2025, and for a system used in sensitive domains. Since Iridium is used to issue medical distress messages with patient information (which raises privacy and safety concerns), to remotely control critical infrastructure, to support government and military communications, or for a variety of private communications, this raises severe confidentiality issues.
%. Figure~\ref{fig:sbd-medical} (redacted) shows an example of a Short Burst Data message we intercepted: it was a medical evacuation report containing patient vitals and location, sent by a team in the field via Iridium SBD. Such information could be highly compromising if intercepted by adversaries (e.g., in conflict zones). Yet, it was plainly visible in our logs.

\begin{figure}
\centering
\includegraphics[width=\columnwidth]{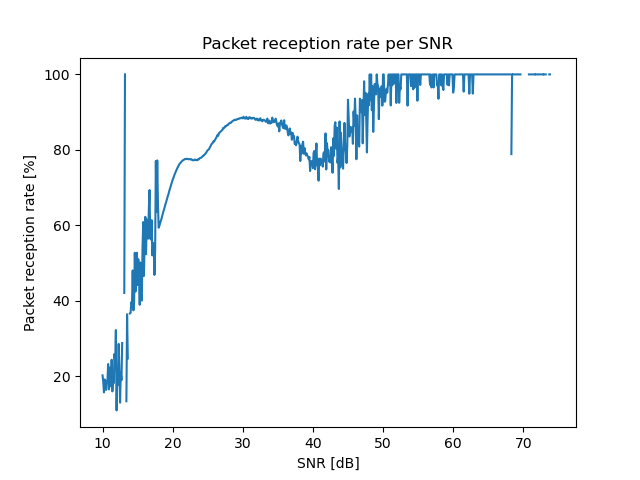}
\caption{Packet reception rate per SNR}
\label{fig:prr-ber}
\end{figure}

From a performance standpoint, our pipeline was able to reassemble fragmented data streams reliably. 
Figure \ref{fig:prr-ber} shows the performance of Iridium message reception. As we can see in the plot we have an area
from 18 dB to around 35 dB SNR, where we have a packet reception rate above 50 \%. 
We can observe a dip in PRR values between 40 dB SNR and 50 dB SNR, which
is caused by the analog-to-digital (ADC) converter of the SDR being saturated, thus distorting some of the bits and lowering the PRR value. This effect tends to disappear for dB value higher than 50, although only very few messages are received in that range.

We evaluated our reassembly success on a sample of 100 Iridium IP sessions that spanned multiple frames. We were able to reconstruct complete higher-layer payloads for 78 of them. The remaining either were incomplete due to capture gaps or the session ended abruptly. In those complete cases, we checked data integrity via known structures and all looked consistent, confirming our reassembly approach’s validity.

%In conclusion, our eavesdropping results demonstrate that an attacker can practically gather intelligence from Iridium traffic on a large scale. We essentially performed a form of SIGINT on Iridium and uncovered private and sensitive data with ease. Considering Iridium’s use in military, maritime, and emergency sectors, the implications are that adversaries could be routinely listening in on these communications if they desired, unless users take extra measures (which, evidence suggests, most do not).

\subsection{Location Tracking and Privacy}
Using the Ring Alert (paging) messages, the RECORD attack \cite{jedermann-record-2024} and recent improvements \cite{liu2025mind} can passively pinpoint user terminal's positions within a few $km^2$. With longer observation times (multiple satellite passes), the localization improves.

In addition, as briefly mentioned in a presentation by the authors of gr-iridium~\cite{ccc-iridium}, there is further location leakage within the metadata of the network as provided by \textit{gr-iridium}. Some of the messages that can be reassembled are part of a GSM-syle protocol that handles the connection to the Iridium network. Probably to speed up the channel acquisition, the mobile terminal sends a Doppler shift, propagation delay and its last known position. The network will in turn respond with the estimated current position of the device. Both communications are done in the clear.

\begin{figure}[t!]
        \centering
        \begin{minipage}{1\columnwidth}
            \centering
            \includegraphics[width=1\textwidth]{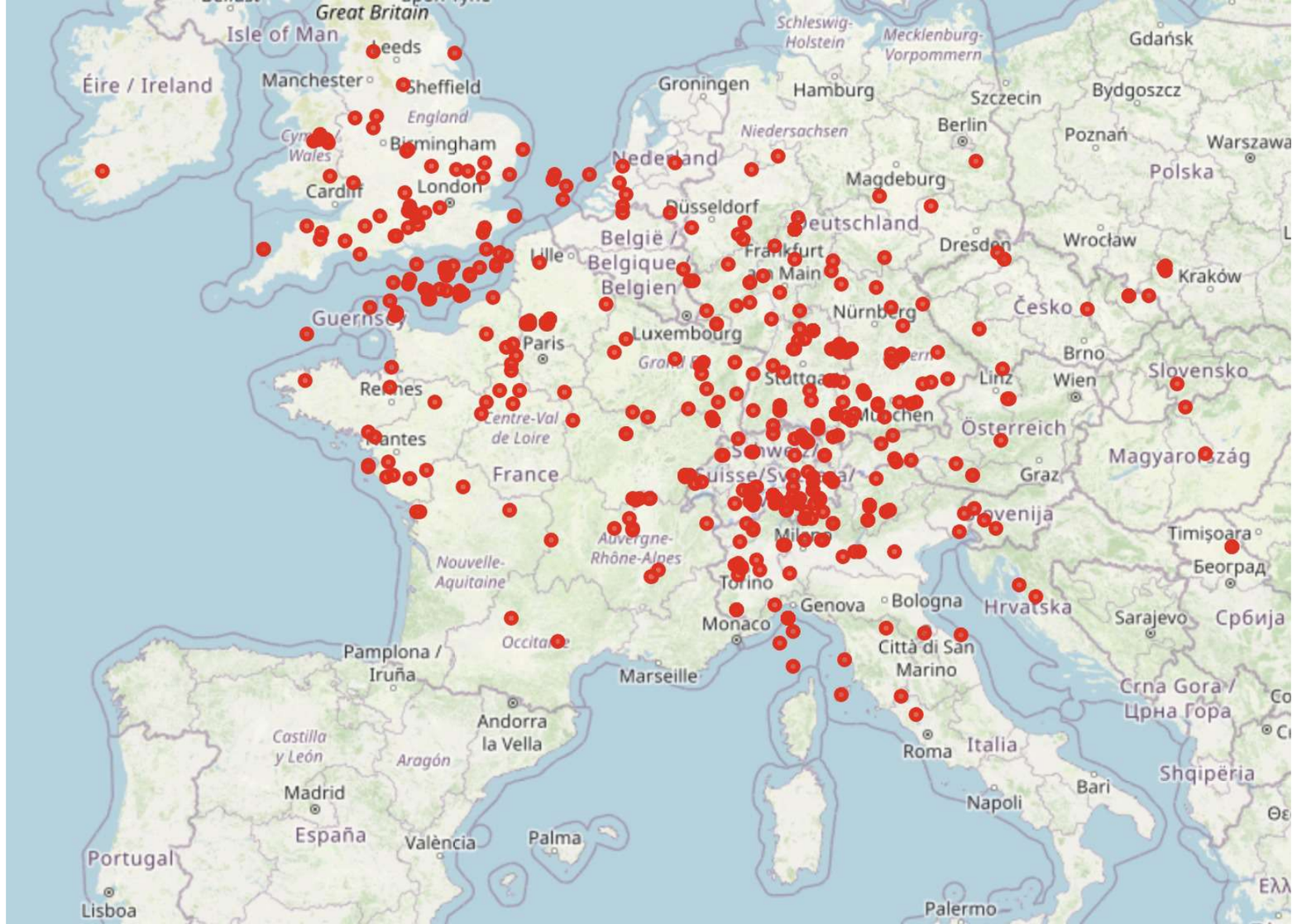}
        \end{minipage}
        \caption{Graph of the 500 recorded estimated coordinates most sent over the air during a period of a week.}
        \label{fig:iridium_map}
\end{figure}

These coordinates are in a geocentric format with a precision to the kilometer on each coordinate. The accuracy of the combined coordinates is thus estimated to be around four kilometers. Figure \ref{fig:iridium_map} shows the 500 coordinates most often recorded over a week, after conversion and projection to a map. The full number of recorded messages that include coordinates in the span of a week is over 130'000, highlighting a serious location privacy leak.

Notably, without analyzing the content of the communication, there is no way to match specific coordinates to a certain terminal. Still, the last position and estimated position pair could also be used to locate a target given some prior information about them or if the locations are unique enough.  

Thus, both through content interception and metadata Iridium leaks user presence and location. The lack of network authentication and use of static beam IDs in messages allow such tracking at scale.

\subsection{Spoofing and Network Impersonation}

\begin{figure}
    \centering
    \includegraphics[width=\linewidth]{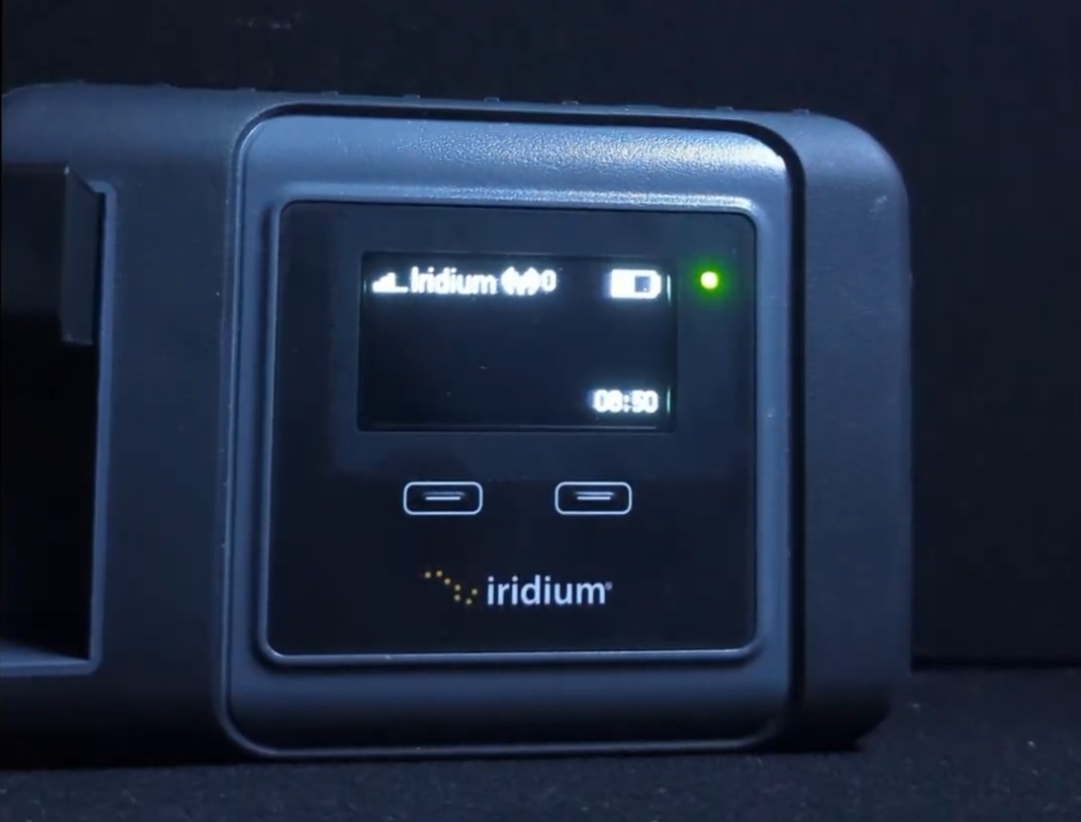}
    \caption{Spoofed Iridium GO!}
    \label{fig:spoofed-iridium-go}
\end{figure}

In our experiments with \textit{gr-iridiumtx}, the attacker is successful in spoofing an Iridium GO! device: the device connected to the attacker signal and authenticated itself against the attacker. The device was then able to acquire the time from the spoofed signal as shown in Figure \ref{fig:spoofed-iridium-go}. Moreover, we observed with Iridium GO! that if the device authenticated against the Iridium network, lost the connection with the Iridium satellite, and was not restarted, we could send a series of Iridium Ring Alerts with the attacker setup. This results in the Iridium device thinking that it is connected to the real Iridium satellite. 

Our attack using \textit{gr-iridiumtx} shows that legacy Iridium devices can be trivially spoofed. This spoofing attack can impact devices used in asset tracking, such as in logistics or transportation; spoofing could lead to inaccurate reporting of the asset's location, potentially causing confusion, delays, or even theft. In emergency situations where Iridium devices are used for SOS signaling and location sharing, a spoofed location could misdirect rescue teams, leading to critical delays in response. Furthermore, if an attacker could successfully spoof the communication channel itself, it could theoretically be possible to send false messages from a compromised device or manipulate data transmissions.

\subsection{Jamming Efficacy}

\begin{figure}
    \centering
    \includegraphics[width=1\linewidth]{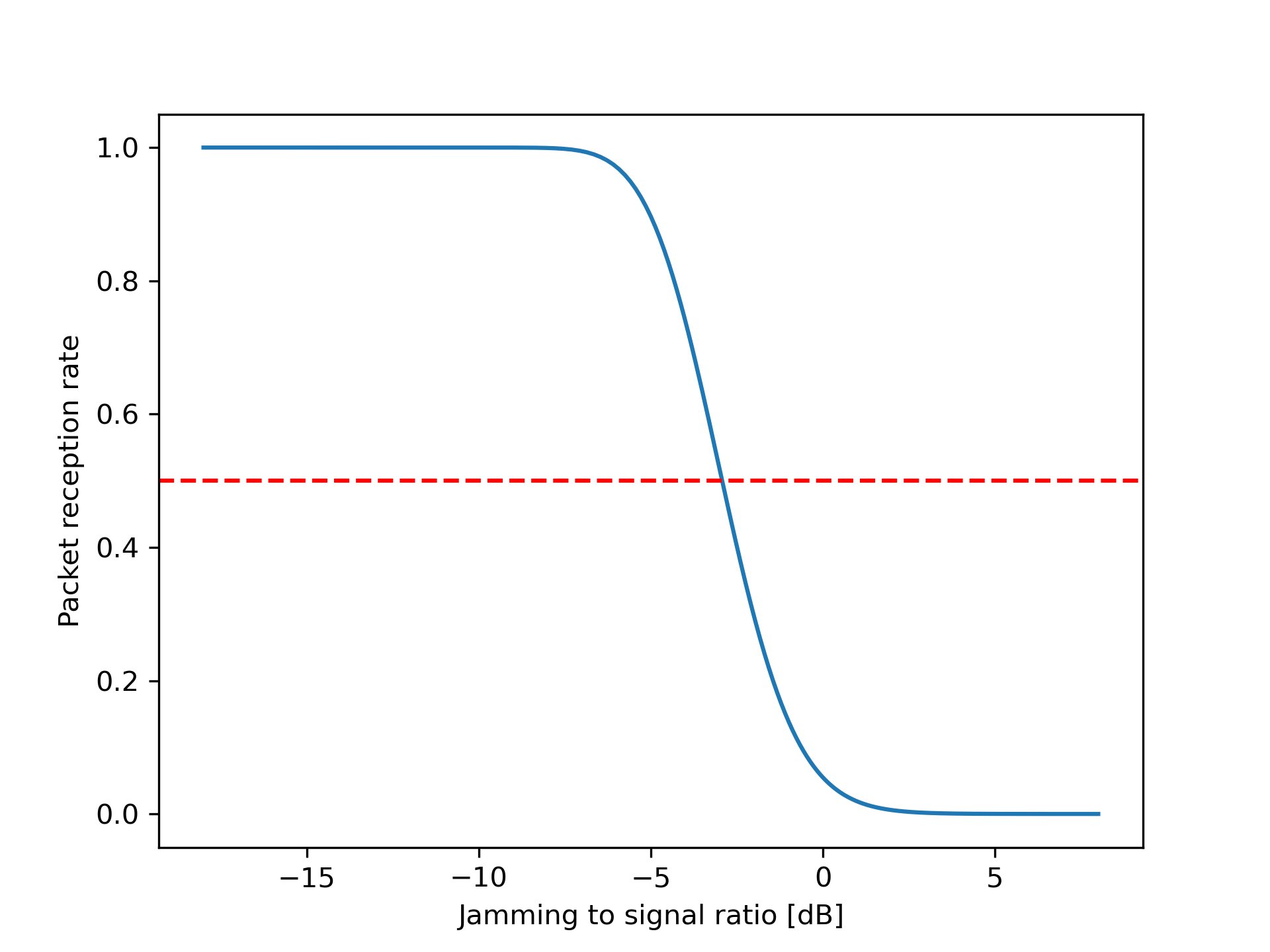}
    \caption{Packet reception rate to the jamming to signal ratio for Iridium Ring Alerts}
    \label{fig:jamming-plot}
\end{figure}

To establish the required signal power budget for a successful jamming attack on Iridium Ring alerts, we follow \cite{smailes-spacesec2024}. In their paper, Smailes et al. derive the value of attacker vs. victim power required to be able to jam Iridium Ring Alerts. From the calculations, we extract the equations to calculate the relation between packet reception rate (PRR) and jamming to signal ratio (J/S) through the intermediary of the bit error rate $p=BER$ and the block error rate $  \mathbb{P}(block \text{ } error)$ as follows:
\begin{equation} \label{eq:ber}
    p = BER = \frac{1}{2}erfc(\sqrt{E_b/N_0}) = \frac{1}{2}erfc(\sqrt{\frac{1}{2J/S}})
\end{equation}

\begin{align}
\mathbb{P}(\text{block error}) 
&= 1 - (1-p)^{31} - 31p(1-p)^{30} \notag \\
&\quad - \binom{31}{2} p^2(1-p)^{29}
\end{align}

\begin{equation} \label{eq:message-error}
    PRR = (1 - \mathbb{P}(block \text{ } error) )^3
\end{equation}
%This relation can be seen in Equations \ref{eq:ber}, \ref{eq:block-error} and \ref{eq:message-error}. 

This relation is visualized in Figure \ref{fig:jamming-plot}. From the plot we can see that at -2.93 dB jamming to signal ratio, the victim would be able to receive only 50\% of the Iridium Ring Alerts. This signifies that at any stronger jamming signal, an attacker should in theory be able to jam the Ring Alerts and disrupt the entire communication for the Iridium receiver.

Our jamming experiments confirm the theory and show that Iridium communication can be fairly easily blocked by overpowering only the Ring Alert messages.
Any jamming attacks that were run with amplitude set to 0.01 would fail.
By increasing the gain of the generated Gaussian noise of the attacker, we were able to jam the Iridium GO! device with 0.05 amplitude on noise source block and HackRF configured to 0 RF gain.  While active, this attack can thus successfully stop Iridium devices from establishing a connection with an Iridium satellite. It affects both legacy and Iridium NEXT devices at the same J/S level, as the channel acquisition protocol is designed around the same principle of acquiring the Iridium Ring Alert first. 

On the downlink, since Ring Alert messages arrive at around $-120$ dBm at a receiver on the Earth's surface and Iridium receivers have limited link margin, a jammer transmitting at a moderate 0 dBm (1 mW) can effectively disrupt the reception of large geographical regions. 
Depending on the strength of the attack, the devices that might be less affected by this attack are devices that only use Iridium Satellite Time and Location (STL) signals for reception of information. The STL is also on the same simplex band as the Iridium Ring Alerts, however, in theory it should be more resilient to simple jamming attacks due to its encoding and correlation construction. This jamming attack on Iridium devices can effectively render any Iridium communication useless within the affected area. This disruption can be particularly critical in emergency situations where people rely on Iridium devices to contact help. It would also block any communication in military situations in remote areas, where Iridium is used as the main asset for critical communication.

\section{Discussion}
Our findings reveal a systemic security problem in the Iridium satellite network: it has inherited and retained the radio link security weaknesses of early GSM technology, despite being used in scenarios requiring high security. We discuss the implications of these results, why they may have persisted, and how they could be addressed going forward.

\subsection{Implications for Users and Infrastructure}
The fact that Iridium traffic is unencrypted means that anyone relying on Iridium for sensitive communications (whether military, humanitarian, or personal) is at risk of eavesdropping. For military and government users, Iridium phones and data links are often used as backups or in remote areas (for instance, Special Forces units or diplomats might carry Iridium phones) \cite{iridiumdefense}. Unless they employ external encryption devices (which, as discussed, is not common), their conversations and messages can be monitored by adversaries. This could lead to intelligence leaks, exposure of operations, or compromise of personal safety. Even the routine transmission of location check-ins or logistic info over Iridium (which we observed) can be enough to tip off an adversary about activities.

For civilian critical infrastructure, Iridium is used in SCADA for pipelines, power grids in remote areas, and more. The lack of encryption or authentication means false commands could potentially be injected if an attacker studied the SCADA protocols. 

The authentication vulnerability is particularly worrying for billing and abuse. A cloned Iridium SIM could be used by an attacker to make calls or use data at the victim’s expense (Iridium service is very expensive).

In summary, we would strongly advise Iridium user communities: if you must use Iridium for something sensitive, use separate end-to-end encryption and assume your Iridium phone calls are listened in on at all times.

\subsection{Design Flaws in Satellite-GSM Integration}
It is worth reflecting on why Iridium ended up in this state. When Iridium was first launched (late 90s), GSM security had not yet been widely broken (COMP128-1 was first cracked around the same time, 1998) and development times were long. Iridium likely adopted GSM authentication and omitted encryption to simplify the system and comply with export regulations (encryption was heavily regulated in the 90s). Satellite links also have higher latency and initially low data rates, perhaps discouraging adding the overhead of encryption. The assumption might have been that the specialized equipment needed to intercept L-band signals and decode them was beyond the reach of most threats at the time. Indeed, SDRs were not commonplace, and a large dish plus analog receivers would be needed to sniff Iridium. This ``security by obscurity'' (relying on physical difficulty and limited public knowledge) may have been deemed sufficient as in many legacy industries. However, as our results show, that assumption is no longer valid and technology has democratized satellite monitoring.

Another oversight is continuing to use COMP128-1 long after it was known to be insecure. GSM operators phased out COMP128-1 in favor of COMP128-2/3 or MILENAGE algorithms in the early 2000s. Iridium perhaps did not update because it would require replacing all SIM cards and updating network infrastructure (which is harder for a satellite system in orbit). Importantly, backward compatibility with old devices may have been prioritized. After all, Iridium NEXT launched around 2017 and still supports legacy devices.

The lack of mutual authentication in Iridium is directly inherited from GSM’s A3/A8 design. GSM later introduced measures in 3G to have the network authenticate to the device (Authentication and Key Agreement protocol). 

\subsection{Towards Securing Iridium}
The development and launch of Iridium NEXT offered an opportunity to enhance security. From the limited public information available, we can assume that mutual authentication and possibly a stronger cipher have been introduced for some higher-layer services \cite{Maurer2022SecurityDigitalAero}. But the retention of backward compatibility for the radio link means legacy devices continue with insecure protocols and radio link vulnerabilities highlighted in the work remain. The number of legacy devices is probably hundreds of thousands still in use, given the number of subscribers in 2019 (1 300 000 subscribers) \cite{iridium-investor-report-2019} and 2025 (2 483 000 subscribers) \cite{iridium-investor-report-2025}. 

Our demonstrated spoofing and jamming attacks are harder to solve purely by protocol changes. To counter spoofing, mutual authentication is needed so that devices can verify a message is from a genuine satellite. This likely requires cryptographic signatures on broadcast messages and the use of Authentication and Key Agreement procedures for control channels. 

Jamming is a physical-layer vulnerability that all wireless systems must deal with. However, satellite constellations could make it harder by frequency hopping or spread spectrum techniques used for example by the more modern Starlink constellation successfully deployed in the war in Ukraine \cite{mittr2025starlinkrepair}. As the original Iridium system is FDMA/TDMA with fixed channels, jamming just one channel is highly effective as we have shown.

\section{Related Work}
Our work builds on and extends findings from both the satellite and GSM security research communities.\\

\textbf{GSM and COMP128 Attacks:} The vulnerabilities of COMP128-1 were first exposed by Briceno, Goldberg, and Wagner in 1998, who obtained and published the algorithm \cite{briceno1998gsm}. Within 8 hours, they were able to reconstruct the secret key using 150,000 requests at a rate of 6.25 requests per second. Subsequent research by Rao et al.~\cite{rao2002partitioning} highlighted how a 128-bit COMP128-1 key could be recovered with as few as 8 chosen plaintexts, exploiting partitioning attacks. As Comp128-1’s weaknesses and the ability to clone GSM SIMs were demonstrated widely around the 2000s, GSM providers moved to COMP128-2/3 and later MILENAGE. To our knowledge, this work is the first to explicitly confirm and examine Iridium’s ongoing use of COMP128-1.\\

\textbf{Prior Iridium Reverse Engineering:} The Iridium protocol was largely reverse-engineered by enthusiast groups and researchers in the 2010s. Notably, Schneider and Zehl presented analyses of Iridium’s air interface at various hacker conferences \cite{ccc-iridium}. They developed the gr-iridium tools \cite{schneider2022gr} for decoding Iridium pager messages and voice, revealing the lack of encryption. Jedermann et al.~recently exploited Iridium’s unencrypted Ring Alerts in their ``RECORD'' attack \cite{jedermann-record-2024} to geolocate users by passively listening to Iridium ring messages over time. However, our work is the first to systematically analyze radio link vulnerabilites in Iridium.\\

\textbf{Satellite Spoofing and Jamming Studies:} The threat of GNSS (GPS) spoofing is well-documented nowadays and exploited widely in the real world, but communications satellites have also been spoofed. Bisping et al.~\cite{bisping-usenix2024} investigate wireless signal spoofing attacks on VSAT communications. Salkield et al. \cite{salkield-wisec2023} systematically examined downlink overshadowing attacks on various satellites. They outline requirements to spoof GEO satellite downlinks and show it’s feasible to overshadow VSAT signals. Iridium, being LEO and spot-beam, actually requires less power to spoof since the attacker can be much closer to the target than the satellite is. Our work confirms Salkield’s general premise in the specific case of Iridium, demonstrating a real overshadow of Iridium signals. On the jamming side, Smailes et al.~and Oligeri et al. introduced techniques to fingerprint Iridium devices so that legitimate signals could be distinguished under interference \cite{smailes-ccs2023,oligeri2022past}. In a follow-up \cite{smailes-spacesec2024} Smailes et al.~quantify the jammer power threshold for disrupting Iridium signals and our empirical results align with their predictions.\\

\textbf{Other Satellite Networks:} Iridium is not alone in facing security issues. Pavur et al.~\cite{pavur-sp2020} showed that eavesdropping on commercial Ku-band VSAT networks used by ships and aircraft was possible, demonstrating the existence of sensitive unencrypted data. Willbold et al.~\cite{willbold-wisec2024} similarly found that current VSAT systems often lack proper encryption or are vulnerable to insider attack, highlighting the industry’s precarious state. Several denial of service attacks on the routing layer have been proposed for modern mega-constellations such as Starlink, for example the ICARUS attack \cite{giuliari2021icarus}. Satellites, in particular small CubeSats, have also been shown at risk of numerous security vulnerabilities \cite{willbold2023space}.

Santamarta \cite{santamarta2} showed that Inmarsat and Iridium ship terminals themselves had backdoors and hardcoded credentials that could be exploited. Those device-level flaws are orthogonal but related; we show that even without backdoors, the core protocol itself is weak.

\section{Conclusion}
In this paper, we presented a thorough security evaluation of the Iridium satellite communication system and found it critically lacking by modern standards despite its widespread usage. We demonstrated that Iridium’s authentication still relies on the broken COMP128-1 algorithm, allowing attackers to recover secret keys and clone SIM cards with ease. We further showed that Iridium provides no encryption on its user link for legacy services, leading to widespread exposure of sensitive information to any eavesdropper. Through both passive analysis and active experiments, we confirmed that adversaries can intercept calls, messages, and data, spoof network signals to mislead devices, and jam or block Iridium communications in a localized area.

Our findings carry significant implications. Any assumption that Iridium communications are secure or private is false – in fact, they are as vulnerable as early-1990s GSM phones, only the signals come from space. This affects thousands of users spanning emergency responders, journalists in conflict zones, sailors, pilots, and militaries. As satellite communication becomes ever more integrated into critical infrastructure and global connectivity, these security issues must be addressed with urgency.

We hope this paper serves as a wake-up call for the satellite communications industry much like the late-90s revelations did for GSM. The technological landscape has changed – cheap SDRs and open-source tools turn yesterday’s theoretical attacks into today’s practical threats. Iridium and similar networks must not ignore known weaknesses, and users should demand better. As we move toward an era of ubiquitous global coverage with thousands of satellites, security cannot be an afterthought. 

\appendix

\section{Ethical Considerations}

Our research was undertaken with the primary goal of improving the security of the Iridium satellite system and related constellations.\\

\textbf{Active experiments.} All experiments involving signal transmissions (e.g., jamming and spoofing) were conducted strictly in compliance with local regulations, and only within controlled environments such as Faraday cages or anechoic chambers. The targets of these experiments were exclusively our own Iridium equipment. These precautions ensured that no signals were transmitted into the wild, and thus no Iridium subscribers or the live Iridium constellation were ever affected. Our SIM cloning experiments were all performed on our own SIM cards that we bought and our test calls with cloned SIM cards were charged to our own subscriptions. Thus, no one experienced a financial loss. \\

\textbf{Passive experiments.} All experiments involving signal reception were performed solely on our own devices or under carefully designed safeguards. We did not attempt to circumvent security mechanisms on devices we did not own, nor did we attempt to decrypt encrypted or scrambled communications. Our analysis was limited to wireless downlink signals broadcast openly by Iridium satellites over wide areas. We decoded these signals only to study their structural and protocol-level properties. If decoded signals revealed protocols that could contain sensitive or personally identifiable information, we immediately stopped the collection of data associated with those protocols. To further mitigate risk, we did not store the payloads of received bursts; instead, we stored only statistics and entropy values, ensuring that reconstruction of the original content is impossible.\\

\textbf{Risk–benefit balance.} We believe the benefits of publicly disclosing the identified radio link vulnerabilities outweigh potential drawbacks. Similar weaknesses in Iridium have previously been noted by white-hat researchers, and given the long-standing capabilities of nation-state adversaries in satellite interception (e.g., \cite{stratign}), it is highly plausible that well-resourced attackers were already aware of these issues. Our contribution is to provide the broader community with a systematic assessment of the threats, thereby raising awareness among stakeholders who may not have previously appreciated their severity.\\

\textbf{Responsible disclosure.} We followed a responsible disclosure process. Iridium Communications Inc. was initially informed of our preliminary findings and tests in December 2023. On August 14, 2025, we submitted a more detailed report of all six presented attacks in this paper through Iridium’s official vulnerability disclosure program. After receiving no response twice, we followed up on August 22, 2025. If Iridium continues not to engage, we intend to notify national cyber security institution such as CISA to enable a coordinated disclosure process and obtain CVEs prior publication at Usenix Security 2026.

\section{Open Science}
In accordance with the Open Science guidelines, we provide all artifacts, including the SIM key extraction, Iridium reception and analysis pipeline, the \textit{gr-iridiumtx} tool, and the Iridium jamming and relay experiments. The artifacts for each of the mentioned parts are separated in respective subfolders at: \url{https://anonymous.4open.science/r/artifacts-security-iridium-B038/}.\\

\textbf{Key Extraction} The artifacts of the SIM key extraction are multiple screenshots, showing the execution of Woron Scan with the extracted (pixeled) $K_i$, the comparison of the open source COMP128-1 implementation vs the original SIM (Figure~\ref{fig:key_extraction}) and the comparison of the original SIM vs the cloned SIM. Further it contains both scripts to produce Figure~\ref{fig:key_extraction}: The open source COMP128-1 implementation from~\cite{Munaut2009comp128v1} (but without the extracted $K_i$) and the script to access the SIM via a card reader from~\cite{Welte2011test}.\\

\textbf{gr-iridiumtx} The artifacts for the \textit{gr-iridiumtx} tool contain a GNU Radio Out-Of-Tree module, which is capable of modulating Iridium frames. The main functionality is it can modulate the Iridium frames at any time and frequency offset within the Iridium spectrum, and save these modulated signals into a SigMF file. \\

\textbf{Pipeline} The artifacts for the data processing pipeline contain Python scripts to run the privacy-oriented pipeline, which extracts metadata from Iridium traffic. In addition, it contains the code to display the statistics of the collected data.\\

\textbf{Jamming and Relay} The artifacts for the Iridium jamming and relay experiments contain the GNU Radio flowgraphs that transmit via HackRF One. One contains a Gaussian jamming flowgraph within the Iridium spectrum, and the other contains the flowgraph that transmits the SigMF files.\\

\bibliographystyle{plain}
\bibliography{refs}

\end{document}